\documentclass[a4paper,11pt]{article}
\pdfoutput=1 

\usepackage{jinstpub} 

\usepackage{array,multirow} 
\usepackage[colorinlistoftodos]{todonotes}

\title{Silica Raman scattering probe for absolute calibration of Thomson scattering spectrometers}

\author[1]{S. Ghazaryan,\note{Corresponding author.}}
\author{M. Kaloyan,}
\author{and C. Niemann}

\affiliation{Department of Physics and Astronomy, University of California Los Angeles\\1040 Veteran Ave., Los Angeles, CA 90095, USA}

\emailAdd{sghazaryan@physics.ucla.edu}

\abstract{We have developed a solid state probe for an absolute irradiance calibration of the Thomson scattering system on the Large Plasma Device (LAPD), based on Raman scattering off silica.
Measurements performed with a triple-grating spectrometer have investigated the intensities of a pulsed laser beam Raman scattered off crystalline and amorphous silica over a range of temperatures of relevance to the LAPD (299 - 498 K). The data were compared with Rayleigh and Raman scattering intensities in gaseous nitrogen. The measurements show that Raman scattering off quartz allows rapid and accurate alignment and calibration of Thomson scattering systems in plasma physics experiments that cannot be calibrated using conventional methods.
}


\begin{document}
\maketitle
\flushbottom

\section{Introduction}
\label{sec:intro}

Thomson scattering (TS) is a reliable model-independent, non-intrusive diagnostic tool in the studies of plasma scattering of electromagnetic radiation. Photons, provided by a pulsed laser beam, scatter off free electrons, which allows characterization of plasma density and temperature with high spatial and temporal resolution  \cite{froula12, carbone14}. Absolute calibration of the detection system and measurement conditions {\it in situ} is essential to avoiding systematic errors. This is typically accomplished via Rayleigh \cite{berni1996} or Raman scattering \cite{flora1987} off gas. Rayleigh scattering is the elastic scattering off electrons bound to heavy particles that produces a signal at the laser wavelength \cite{iordanova12}. Therefore, stray light reduction measures must be taken to distinguish the Rayleigh signal from stray light \cite {Kempkens00}. Raman scattering is characterized by specific frequency changes of incident photons as a result of a transitions in rotational and vibrational states of the molecule \cite{vries08}. The Raman signals appear on both sides of the photon wavelength, separate from the stray light, but within the TS spectral range  \cite{toshihiko84}.  Previous studies examined calibration of TS with Raman scattering off O$_2$ \cite{gessel12}, H$_2$ \cite{toshihiko84}, and N$_2$ \cite{vries08}. A potential drawback of this approach is that the Raman scattering intensity is less than that of Rayleigh by several orders of magnitude and requires much longer exposure times \cite{bisson05}. As such, using Raman scattering in conjunction with Rayleigh scattering can calibrate the photon collection efficiency with greater precision, while validating that the intensity response of the detector is linear.

A TS diagnostic is currently being developed for the Large Plasma Device (LAPD) \cite{gekelman2016} at the University of California Los Angeles, which can facilitate research on collisionless shocks \cite{niemann14}, magnetic reconnection \cite{gekelman2007}, and mini-magnetospheres \cite{gargate08}. 
This diagnostic will be described in the future but is based on the same spectrometer used for this work.
Given that the LAPD is a user facility, it cannot be filled with gas regularly for alignment and calibration.  We propose an absolute calibration via Raman scattering off a quartz crystal probe. To our knowledge, this is the first time that Raman scattering off quartz and fused silica are being used for TS calibration. This method offers significant advantages over other techniques that require the placement of a calibrated light source into the scattering volume \cite{yamauchi1986}. 
We have cross calibrated crystalline (quartz) and amorphous silica (fused silica) using both Rayleigh and Raman scattering off nitrogen over a range of pressures. Quartz has the advantage of producing distinct and bright Raman lines that simplify calibration, while amorphous silica is orientation independent and produces a photon count comparable to TS.
The Raman spectrum of quartz and line intensities depend on the crystal temperature. During LAPD plasma operation the crystal can heat up to temperatures as high as 200$^o$C, significantly affecting the calibration factor. However, the crystal temperature can be directly deduced from the measured Raman Stokes to anti-Stokes line intensity ratio. 
We present Raman line intensity measurements for crystal temperatures between 299 and 498 K.

\section{Experimental setup}
\label{sec:setup}
\begin{figure}[htbp]
    \centering 
    \includegraphics[width=15.cm]{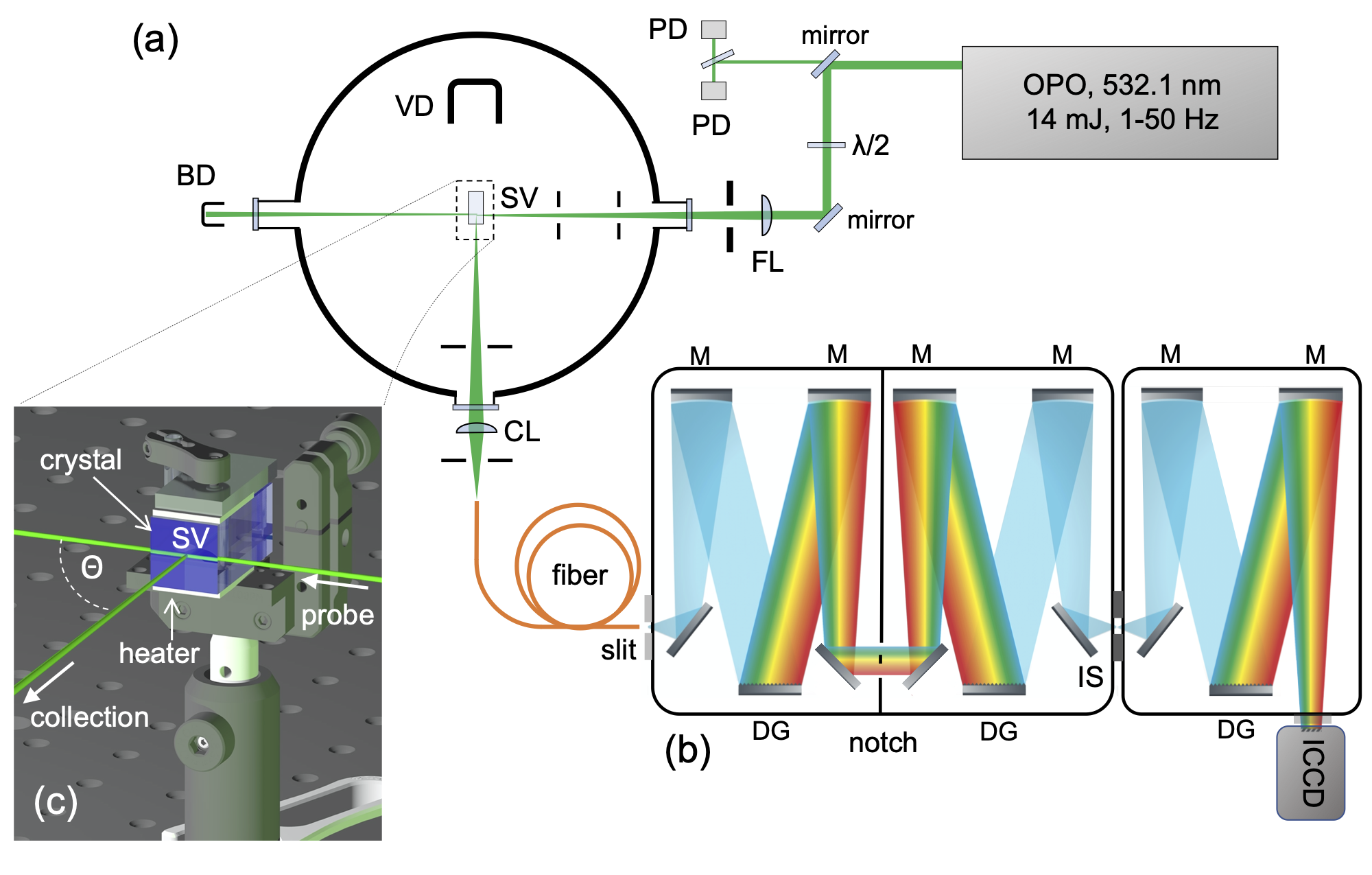}
    \caption{\label{fig:setup} Overview of the experimental setup showing (a) schematic view of the experimental setup, (b) internal arrangement of the spectrometer, and (c) magnified 3D view of the scattering geometry {\bf {a.}} Electromagnetic waves are generated by the optical parametric oscillator (OPO). Mirrors are used to guide the beam to the vacuum chamber. The viewing dump (VD) and baffles inside and outside the chamber reduce stray light contribution. Photodiodes (PD) along the path of the laser are utilized for energy measurement. A focusing lens (FL) is placed in front of the vacuum chamber in order to focus the beam into the scattering volume (SV). Beam is trapped by a beam dump (BD). Scattered light is collected with a collection lens (CL) and focused onto a single fiber, which feeds directly into the spectrometer. {\bf {b.}} Light that enters the spectrometer through the input slit is reflected by mirrors (M) and dispersed by three diffraction gratings (DG) as it is passes through the notch filter and is focused onto the intermediate slit (IS) on its way to the detector (ICCD camera). {\bf {c.}} Crystal is sandwiched between two heaters (white). Beam is directed towards the crystal, scattered, and collected at $\Theta=90^o$ by the collection lens. 
}
\end{figure}

A schematic drawing of the scattering setup is shown in figure \ref{fig:setup}. The laser source employed for our measurements is a diode pumped optical parametric oscillator (OPO), tuned to the second harmonic of Nd:YAG ($\lambda_i=532.1$ nm, E$_{\rm{pulse}}$=10-14 mJ, pulse length $\tau_L$ = 4 ns). The repetition rate was set to 1 Hz to match the parameters that will be used in the LAPD. Raman spectra in static nitrogen were also recorded at 50 Hz. The laser linewidth is 5 cm$^{-1}$ which is slightly less than the spectrometer resolution. The laser emits pulses with shot-to-shot energy fluctuations of 10$\%$. Pulse energy was recorded on every shot by two photodiodes read out via a 16 bit digitizer that monitors a fraction of the laser beam energy leaking through a turning mirror. Intensities of all spectra were post-processed to account for laser energy fluctuations.  The two energy pickups agree within better than 2$\%$. 
%
The laser beam was focused into the 38 $\times$ 19 $\times$ 19~mm$^3$ cuboid crystal mounted in the center of a spherical 75~cm diameter vacuum chamber using a slow f/100 spherical lens with a focal length of f=50 cm. 
This configuration produced a "pencil" beam with a diameter of 0.5~mm full-width at half maximum
(FWHM) over a Rayleigh length of several centimeters, at a fluence around 10~J/cm$^2$. 
The beam entered the vacuum chamber through a flat anti-reflex coated quartz window and was incident on the crystal normal to one of the elongated surfaces as shown in figure \ref{fig:setup}c.

Scattered light was collected in an f/20 cone perpendicular to the probe beam ($\Theta=90^o$) to match what will be used in the LAPD and to maximize scattering probability and spatial resolution \cite{vries08}. 
The laser beam was linearly polarized perpendicular to the direction of the collected light ($\varphi = 90^o$).
A collection lens with a focal length of 10~cm was placed such as to focus the scattered beam at f/5 into a single 200~$\mu m$ optical fiber with a numerical aperture of NA=0.22, coupled directly to the spectrometer input slit. 
The scattering volume is defined by the intersection of the focused laser beam and the projection of the fiber core cross section onto the beam.
In order to reduce alignment sensitivity,
the collection system was designed such that the 4$\times$ magnified fiber projection
exceeds the beam size by a factor of two.
Several measures were taken to achieve high stray light rejection. A viewing dump and baffles were placed inside the scattering chamber around the laser beam and the collection branch. Irises were placed along the path of the beam to filter non-collimated beam components. The scattering volume was produced in the center of the chamber, away from its walls and windows. The beam terminates in a beam trap 100 cm from the scattering volume to minimize stray light via temporal filtering. 
For Rayleigh and Raman scattering measurements in nitrogen
the crystal was removed but the scattering volume stayed the same.
The chamber was backfilled with 99.998$\%$ pure nitrogen gas at room temperature at pressures ranging from 0.03 to 1 standard atmospheres.  The gas was allowed to settle after every pressure change and the pressure was monitored by a capacitance manometer.

Since the Raman scattering signal is several orders of magnitude dimmer than the unshifted Rayleigh signal and stray light, a triple-grating spectrometer (TGS) was used that enhances the performance of the diagnostic by increased stray light rejection. 
Specifically, a triple grating f/4 Czerney-Turner type imaging spectrometer with an 0.5 m focal length and three 1200 grooves/mm holographic gratings was used. 
As illustrated in figure \ref{fig:setup}b, 
scattered light from the fiber entering the spectrometer through the input slit is collimated and guided through a notch filter and intermediate slit by a number of internal mirrors before it reaches the detector. The first two gratings are used in subtractive dispersion to reject more stray light. The last grating disperses the light onto the detector. Toroidal mirrors were used to image the input slit onto the intermediate slit and detector.
The TGS was equipped with a notch mask 
between the double-subtractive spectrometers to block wavelengths around 532 nm.
We used 0.75 mm, or 1 mm wide and 50 $\mu$m thick stainless steel notch masks as needed, corresponding to blocking widths of 1.5 nm and 2 nm, respectively. 
The spectral resolution was measured to be 0.21 nm for a 100 $\mu$m input slit and a 1.0 mm intermediate slit, with a full wavelength coverage of 19.4 nm. 
The total transmission through the spectrometer was measured to be 25$\%$ and is mostly determined by the reflectivity of the three aluminum-coated holographic gratings blazed at 500 nm. 
%


Spectra were recorded on an  image intensified charge-coupled device (ICCD) camera, directly mounted to the output of the third spectrometer stage. The camera was equipped with a  {\it Generation III} photocathode with 50$\%$ quantum efficiency at 532 nm.
All spectra were recorded at maximum micro-channel plate gain of 233, and with an exposure time of 10 ns, with the 4 ns beam centered in the 10 ns window to account for jitter.
In order to increase the pixel count 
a 2$\times$2 pixel hardware binning was applied to all images. The spectra were further software binned over the entire fiber and slit vertically into 512 total horizontal bins with 0.038 nm/bin. 
An equal number of data and background shots were recorded back to back. 
The 20 e$^-$ readout noise of the ICCD cooled to -20$^o$C is negligible when binning the spectrum and averaging multiple shots. In the absence of plasma fluorescence and since ambient light was eliminated by the short exposure time, the noise was determined solely by the shot noise of the laser. The signal to noise (SNR) therefore scaled as SNR $\sim \sqrt{N}$, where $N$ is the number of laser shots. We typically averaged 100 laser shots and 100 background shots at 1 Hz for each Rayleigh spectrum, while the Raman spectra in nitrogen were averaged over 25,000 laser shots and an equal number of background shots to increase SNR.

We investigated Raman scattering off synthetic crystal quartz and fused silica cuboids of identical dimensions. Optical surfaces for laser entry and exit and scattered light collection were polished at $\lambda$/10 and 20-10 scratch-dig, while the other three surfaces were frosted. The crystals' surfaces were aligned to be normal to the beam within $\pm 1 ^o$ using the laser back-reflection. Refraction at the crystal interface is negligible as long as alignment is kept within reasonable limits. For example, yawing the crystal by as large an angle as $\pm$5$^o$ only displaces the beam in the center of the cuboid by $\pm$0.2 mm, which is less than the diameter of the scattering source and does not affect the collection efficiency. This has been confirmed in the measurements below.
The optical axis of the crystal was aligned parallel to the collection axis. Quartz and fused silica produce Raman emission within the spectral range of a typical Thomson scattering spectrometer and have a high optical damage threshold \cite{said1995} in excess of 100 J/cm$^2$ so that they can be placed in the focus of the pulsed probe beam. In this experiment the crystals were mounted on a kinematic stage. In the LAPD they will be inserted on a standard probe shaft using a motorized stage \cite{gekelman1991} via a vacuum interlock for rapid TS alignment and calibration.
In order to investigate the Raman spectra as a function of temperature the quartz crystal was sandwiched between two 25 W ceramic heaters and the temperature was stabilized with a thermocouple and proportional-integral-derivative (PID) controller. The maximum temperature was limited to 225$^o$C by heat conduction in the ambient air and the ceramic optical post and the wire leads.

\section{Results and discussion}
\subsection{Notch filter}
\label{sec:notch}
\begin{figure}[htbp]
    \centering 
    \includegraphics[width=15cm]{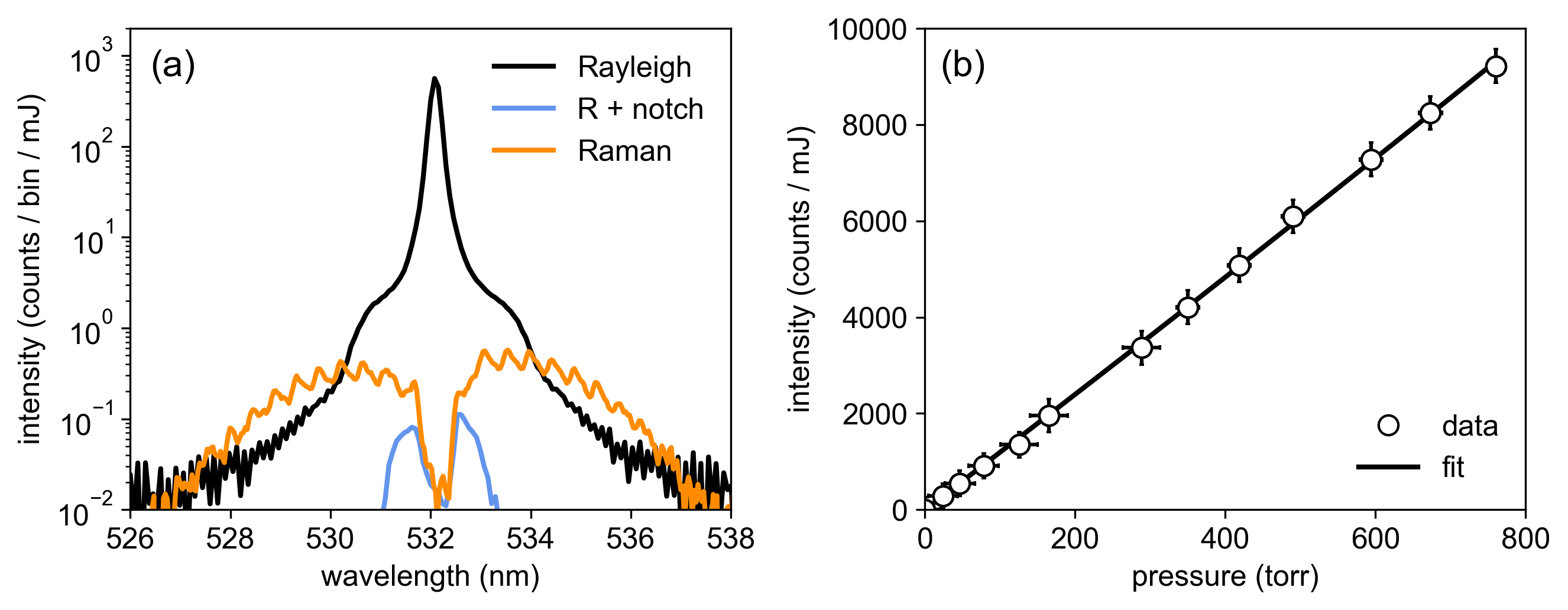}
    \caption{\label{fig:rayleigh} (a) Comparison of the Rayleigh (black) and Raman scattering spectra (orange) of nitrogen at 750 torr for a 100 $\mu$m slit. Since the instrument broadened Rayleigh signal cannot be separated from the Raman spectrum we actually plot the stray light scaled to the Rayleigh scattering amplitude, which also represents the instrument function.
    Without the notch, the broad Lorentzian wings of the Rayleigh line overpower the much fainter Raman spectrum. The 1.5 nm notch reduces the central 532 nm wavelength by five orders of magnitude and the wings by three (blue line), and reveal the Raman spectrum.  (b) The measured Rayleigh scattering intensity in N$_2$ increases linearly with pressure as expected. The absence of a signal at zero pressure shows that stray light was effectively eliminated in this experiment.}
\end{figure}

Figure \ref{fig:rayleigh}a shows the instrument function for a 100 $\mu$m slit and without the notch filter, measured by scattering the probe-laser off a frosted glass diffuser. The peak amplitude was scaled to the Rayleigh scattering signal amplitude in nitrogen at 750 torr, to produce the equivalent Rayleigh scattering profile without the Raman spectrum. The instrument profile is the convolution of the Gaussian contribution due to the aberrated slit width, and the Lorentzian profile caused by diffraction off the three gratings. A Raman scattered spectrum recorded at identical conditions with the notch filter is shown for comparison (orange line) and is three orders of magnitude smaller in amplitude than the Rayleigh signal. Without a notch filter, the broad Lorentzian wings of the instrument broadened Rayleigh signal overpower the much fainter Raman signal. The 1.5 nm width notch reduces the Rayleigh power at 532 nm by five orders of magnitude (blue curve) and the wings by three, and makes the detection of the faint Raman signal possible. The notch reduces the intensity of the wings outside its 1.5 nm width, since it removes 532 nm light before it diffracts off the final two gratings.
The total extinction of the 532 nm instrument broadened line was 2$\cdot 10^4$, equivalent to an optical density of 4.3. The 2 mm wide notch in combination with a 1 mm intermediate slit had an optical density of 5.0.

\subsection{Absolute spectrometer calibration}
\label{sec:calibration}

For light scattering off microscopic particles, and when the scattering cross section depends on the scattering direction, as is the case for Thomson, Rayleigh, or Raman scattering, the scattered power $P_s$ is 
\begin{equation}
    P_s = P_i \cdot n\cdot L_{det} \cdot {\frac{d\sigma}{d\Omega}}\cdot \Delta\Omega\ ,
\end{equation}
where $P_i$ is the incident power, $n$ is the density of scattering particles, $L_{det}$ is the length of the scattering volume, $d\sigma/d\Omega$ is the differential cross section, and $\Delta\Omega$ is the solid angle of detection. 
The total number of photons emitted per pulse is the total energy divided by the energy of one photon or $E_i/(h\nu_i)$, where $\nu_i$ is the frequency of the incident laser light. However, only a very small fraction of these photons encounter and scatter off the particles. 
Laser parameters can be expressed in terms of the power and intensity of photons in the scattering volume. Since $P_i = E_i/\tau_L$, where $\tau_L$ is the pulse length, and intensity $I_L = P_i/A$ is power per cross sectional area, the number of photons supplied by the light source per pulse can be quantified as ${\tau_L \cdot I_L}/{(h\nu_i)}$. 
The scattering volume $\Delta V =A\cdot L_{det}$, and solid angle of detection, $\Delta\Omega$ are determined by the collection lens. Thus, the total number of photons scattered into the solid angle of the collection lens is \cite{schaeffer10}:

\begin{equation}
N = \frac{\tau_L \cdot I_L}{h\nu_i} \Delta V \Delta \Omega \cdot n\cdot \frac{d \sigma}{d\Omega}\ .
\end{equation}
Transmission efficiency through optical components should also be taken into consideration since only a fraction the photons scattered into the solid angle $\Delta \Omega$ make it to the detector. The optical transmission through the vacuum window, lens, optical fiber, and spectrometer is only around $\mu=0.1$, mostly due to the low reflectivity of the three diffraction gratings in the triple spectrometer. Photons are converted to photo-electrons in the iCCD camera with quantum efficiency $\eta=0.5$, which are multiplied by the MCP of gain $G=233$ before being recorded on the iCCD.
The total number of photo-electrons (counts) on the detector area is then
\begin{equation}
N_{pe} = \underbrace{ \overbrace{ \frac{\tau_L \cdot I_L}{h\nu_i}}^{\rm{laser}}  \overbrace{  \Delta V  \Delta \Omega}^{\rm{collection}} \cdot  \overbrace{ \mu\ \eta \ G}^{\rm{optics}}}_{k}  \cdot\ n \cdot \frac{d\sigma}{d\Omega}\ .
\end{equation}
The experimental throughput parameter $k$ depends on the probe laser parameters, the collection optics and scattering volume, transmission through optical components and the spectrometer, and the camera sensitivity and gain.

For Thomson scattering,  the total number of counts $N_T$ in a measured spectrum is proportional to the electron density $n_e$
\begin{equation}
N_T = k\cdot n_e \frac{d\sigma_T}{d\Omega}\ ,
\label{tscount}
\end{equation}
where the differential cross section is $\frac{d\sigma_T}{d\Omega}=r_e^2$ for scattering perpendicular to the probe beam, and $r_e = 2.818\cdot 10^{-15}$ m is the classical electron radius. If $k$ is known, $n_e$ can be deduced from the measured TS counts. Since $k$ is difficult to calculate accurately, it is typically calibrated {\it in situ}.
This can be accomplished, for example, via Rayleigh or Raman scattering.
The total number of counts in the Rayleigh scattering spectrum is given by \begin{equation}
    \label{eqn:rayleigh}
    N_R = k\cdot n_{\rm{gas}} \frac{d\sigma_R}{d\Omega}\ ,
\end{equation} 
where the differential cross section for nitrogen at 532 nm is $d\sigma_R / d\Omega = 6.07\cdot 10^{-32}\ m^2$
for scattering perpendicular to the probe beam \cite{vds}. 
Using a 100 $\mu$m input slit, we measured (4.7 $\pm$ 0.2) $\cdot 10^4$ total counts at 10 mJ and 750 torr,  corresponding to $k= (3.13\pm 0.14) \cdot 10^{12}\ cm^{-1}$.
Figure \ref{fig:rayleigh}b shows the measured Rayleigh scattering intensity in nitrogen as a function of pressure.  The absence of a signal at zero pressure indicates that stray light was effectively eliminated in this experiment.
The slope of the linear fit is proportional to $k$.

\begin{figure}[htbp]
    \centering 
    \includegraphics[width=15cm]{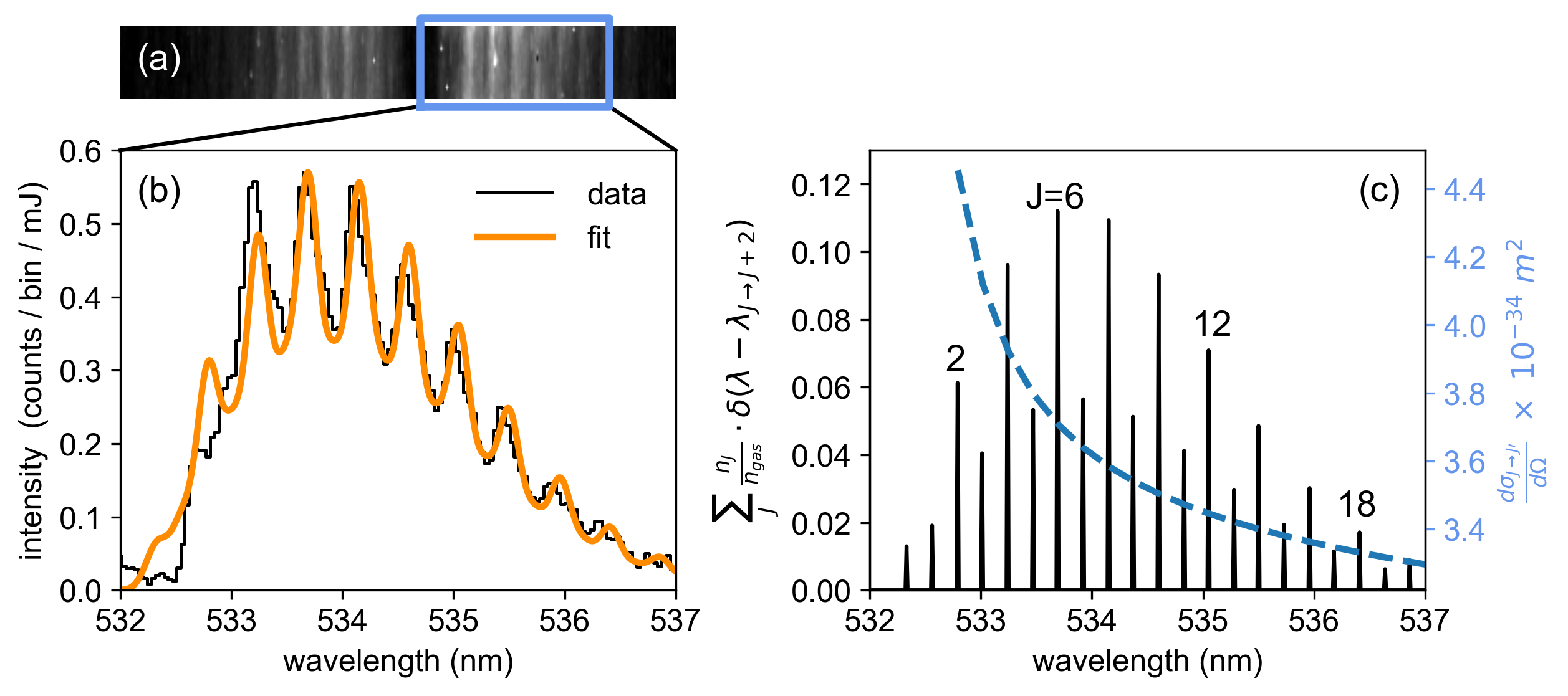}
    \caption{\label{fig:raman} (a) Raw ICCD image showing the Raman spectrum in nitrogen with the laser line blocked by the notch in the center. (b) Measured Stokes Raman spectrum (black) and the best fit of the theoretical spectrum (orange). The total area under the fit is used to determine the measured Raman scattering signal count. (c) Calculated rotational state densities at room temperature and their spectral distribution. The statistical weight factor $g_J$ causes the alternating amplitudes of adjacent transition lines. The differential cross section decreases with wavelength (dashed line).}
\end{figure}

Figure \ref{fig:raman} shows a raw Raman scattering ICCD image averaged over 25,000 laser shots (a) and the corresponding spectrum of the red-shifted Stokes lines (b). 
The Raman spectrum is composed of a large number of narrow peaks on both sides of the laser line, blocked by the notch and shown in the center of figure \ref{fig:raman}a. 
Each line corresponds to a different rovibrational transition from one rotational state $J$ to another $J'$, induced by the inelastic scattering process. Only transitions $J\rightarrow J + 2$ (Stokes) and $J\rightarrow J - 2$ (anti-Stokes) are allowed, where the integer $J$ is the rotational quantum number. 
In the following we will review the analytical expressions needed to fit the rotational Raman spectrum as described in more detail elsewhere \cite{penney1974, deregt1998, vds}.
The wavelength of the red-shifted Stokes Raman lines may be approximated by
\begin{equation}
    \lambda_{J\rightarrow J+2} \approx \lambda_i + \frac{\lambda_i^2}{hc}\cdot B(4J + 6) \ ,
\end{equation}
where the rotational constant is $B=2.48\cdot 10^{-4} eV$ for nitrogen and $hc=1240\ eV\cdot nm$.
The total counts observed for a single Raman line $J$ is given by
\begin{equation}
    N_{J\rightarrow J'} = k\cdot n_J \frac{d\sigma_{J\rightarrow J'}}{d\Omega}\
\end{equation}
where the density $n_J$ of a rotational state at temperature $T$ is determined by the Boltzmann distribution
\begin{equation}
    n_J = n_{\rm{gas}}\frac{g_J(2J+1)}{Q} exp\left(-\frac{E_J}{k_BT}\right).
\end{equation}
Here $g_J$ is a statistical weight factor, which is $g_J = 6$ or $3$ for even or odd $J$, respectively. $Q$ is the partition sum \cite{herzberg1950}
\begin{equation}
Q = \sum_J g_J(2J+1) \cdot exp\left(- \frac{E_J}{k_BT}\right) \approx \frac{9 k_B T}{B}\ ,
\end{equation}
and the energy of a rotational state $J$ is given by 
\begin{equation}
    E_J = B\cdot J(J+1).
\end{equation}
Figure \ref{fig:raman}c shows the densities of the Stokes rotational states visible at room temperature and their spectral distribution. The statistical weight factor causes the alternating amplitudes of adjacent transition lines. 
The differential cross section of an individual transition $d\sigma_{J\rightarrow J'}/d\Omega$ depends on the Placzek-Teller coefficients, and the polarizability anisotropy, which stems from measurements \cite{penney1974} and was interpolated at 532 nm \cite{vds}.
The cross section varies with wavelength as shown in figure \ref{fig:raman}c around a weighted average of 3.8$\cdot 10^{-34} m^2$.
The total number of counts in the brightest Stokes Raman line ($J = 6 \rightarrow 8$) is given by
\begin{equation}
    \label{eqn:raman-calib}
    N_{\rm{6\rightarrow 8}} = \ k\cdot n_{\rm{gas}} \cdot 4.20\cdot 10^{-35} m^2\ .
\end{equation}
At room temperature (T=295 K) only the lowest 20-25 lines are visible on either side of the spectrum and the total number of counts in all visible Stokes lines is given by
\begin{equation}
    N_S = \sum_{J=0}^{25} N_{J\rightarrow J+2} = k\cdot n_{\rm{gas}} \cdot 3.82\cdot 10^{-34} m^2\ .
\end{equation}
Similarly, the total number of counts in the anti-Stokes spectrum is given by
\begin{equation}
    N_{\rm{AS}} = \sum_{J=2}^{27} N_{J\rightarrow J-2} = k\cdot n_{\rm{gas}} \cdot 2.68\cdot 10^{-34} m^2\ ,
\end{equation}
and the total counts in the entire rotational Raman spectrum by
\begin{equation}
    \label{eqn:raman-calib}
    N_{\rm{Raman}} = N_{\rm{AS}} + N_S \ = \ k\cdot n_{\rm{gas}} \cdot 6.49\cdot 10^{-34} m^2\ .
\end{equation}
Given that $k$ remains constant, dividing equation \ref{eqn:rayleigh} by equation \ref{eqn:raman-calib} shows that the total number of counts in the Rayleigh spectrum is 94 times that of the  total counts in the rotational Raman spectrum
\begin{equation}
    \label{rayleigh-raman}
    N_R = 93.7 \cdot N_{Raman}\ ,
\end{equation}
and the ratio between the intensities of Rayleigh and the brightest Raman line ($J=6\rightarrow8$) is
\begin{equation}
    \label{raman68}
    N_R = 1.47\cdot 10^3 \cdot N_{6\rightarrow 8}\ .
\end{equation}
The Rayleigh to Raman amplitude ratio is larger than the total integrated intensity ratio, since the Raman spectrum is much wider.
Similarly, the ratio between the measured counts in the Raman spectrum and in the Thomson spectrum could then be used to deduce the electron density $n_e$ \cite{vds}
\begin{equation}
    \frac{N_{\rm{Raman}}}{N_T} = 8.15\cdot 10^5 \ \cdot \frac{n_{\rm{gas}}}{n_e}\ .
\end{equation}

\noindent
The spectral resolution in this experiment did only allow the separation of the even-J lines while the weaker odd-J lines were buried in the wings. Furthermore, the lines closest to $\lambda_i$ were partially suppressed by the notch. Instead of directly integrating the measured Raman spectrum, the total count was therefore determined from the area under the synthetic fit to the measured spectrum.
The counts in the Raman fine-structure spectrum as a function of wavelength is 
\begin{equation}
    \label{raman}
    I_{\rm{fs}}(\lambda) = k \sum_{J}  n_J \frac{d\sigma_{J\rightarrow J'}}{d\Omega} \delta(\lambda - \lambda_{J\rightarrow J'})\ ,
\end{equation}
where $\delta(\lambda - \lambda_{J\rightarrow J'})$ is the Dirac delta function.
Natural line broadening and pressure broadening are negligible, and the width of the measured Raman peaks is determined solely by the instrument function. The synthetic Raman spectrum $I_{\rm{fit}} = I_{\rm{fs}} \circledast I_{\rm{instr}}$ 
can then be constructed via convolution of the calculated fine structure spectrum $I_{\rm{fs}}$ and the experimentally measured instrument function $I_{\rm{instr}}$.
Given that the width of a bin is the same for the experimental and the synthetic spectrum (0.038 nm/bin) the sum of the synthetic spectrum is equivalent to equation \ref{eqn:raman-calib}.
\begin{equation}
    N_{\rm{Raman}} = \sum_{\lambda} I_{\rm{fit}}\ \ .
\end{equation}
Using a 100 $\mu$m slit we measured a total Raman scattering signal of N$_{\rm{Raman}}$=560$\pm$80 counts at 10 mJ and 750 torr (figure \ref{fig:raman}b).
This measured ratio of total Rayleigh to Raman scattered photons of $N_{\rm R}/N_{\rm{Raman}}$ = 104$\pm 15$ agrees well with equation \ref{rayleigh-raman}, which indicates that the camera response is linear over the dynamic range used. The measured ratio of Rayleigh peak amplitude to J$_{6\rightarrow8}$ Raman peak amplitude is slightly larger than predicted by equation \ref{raman68}, since adjacent lines overlap.

\subsection{Raman scattering off silica}
\label{sec:quartz}
\begin{figure}[!ht]
    \centering 
    \includegraphics[width=11cm]{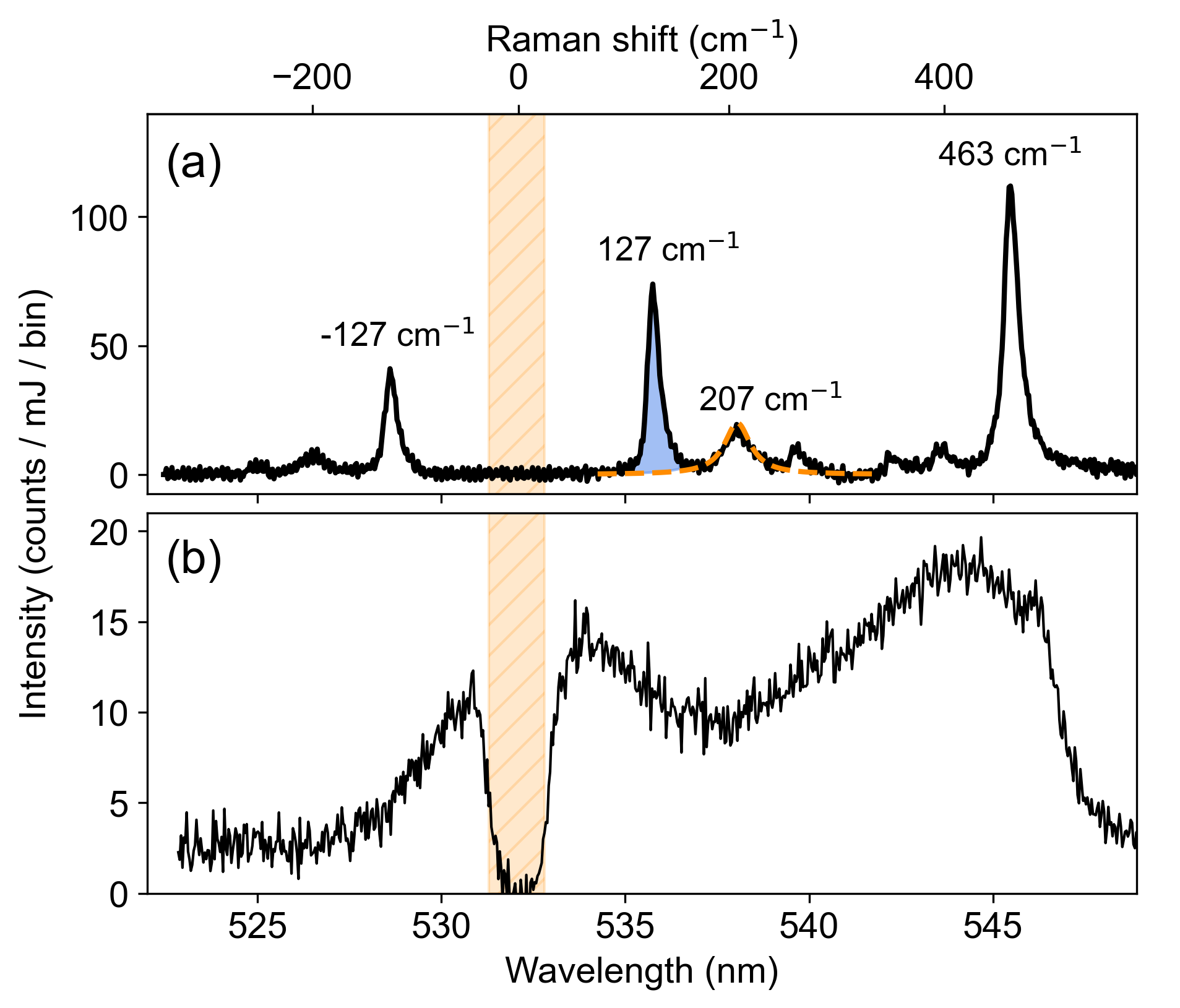}
    \caption{\label{fig:quartz} Raman spectra of crystalline (a) and amorpheous silica (b) obtained with a 2 nm wide notch mask. The area under the 127 cm$^{-1}$ line in quartz is used for calibration (shaded blue), after subtracting the broad Lorentzian wings of the 207 cm$^{-1}$ line (orange dashed line). The spectrum shows both the Stokes and anti-Stokes lines on either side of the notch filter (shaded orange).}
\end{figure}

Figure \ref{fig:quartz} compares the Raman spectra of crystalline (a) and amorphous (b) silica. The data seamlessly combine multiple spectra obtained with different spectrometer indications to extend the spectral range.  
The spectra show both the red-shifted Stokes and blue-shifted anti-Stokes lines on either side of the 532 nm laser line.
Rayleigh scattering peaks have amplitudes several orders of magnitude higher than the Raman spectra, and were blocked by a 2 nm wide notch.
Quartz spectra were integrated over 200 laser shots and 200 background shots, while the fainter spectra of amorphous silica were averaged over 5000 laser shots.
The quartz Raman spectrum shows distinct peaks at 127 cm$^{-1}$, 207 cm$^{-1}$, and 463 cm$^{-1}$, that are slightly broader than the instrument function. The red-shifted Stokes lines are typically used for sample analysis purposes since they are brighter and less temperature dependent than their blue-shifted anti-Stokes counterparts.
We used the 127 cm$^{-1}$ peak as the calibration reference since it is closest to the laser line but outside the notch. It overlaps with the wide Lorentzian wings of the adjacent 207 cm$^{-1}$ line \cite{krishnamurti1958}, which must be subtracted. The strongest peak at 463 cm$^{-1}$, which corresponds to symmetric stretching-bending modes of Si-O-Si \cite{kathleen1994, biswas2018} is outside the spectral range of a typical Thomson scattering diagnostic. 
By contrast, the Raman spectrum of amorphous silica shows a broad band emission with a maximum at around 450 cm$^{-1}$, which is due to the Si-O-Si bond rocking and bending in the SiO$_4$ tetrahedra \cite{mcmillan1984}.
It is interesting to note, that the total integrated intensity in the amorpheous and in the crystalline silica Raman spectra are almost identical. However, in the quartz spectrum intensity is concentrated in a few bright lines, and the  amplitudes are an order of magnitude larger. 
In fused silica the Rayleigh signal is more than two orders of magnitude higher in amplitude than the Raman signal. In quartz the Rayleigh signal amplitude is one order of magnitude higher than the Raman amplitude.
The Rayleigh signal of fused silica is also an order of magnitude brighter than the Rayleigh signal from quartz.

\begin{figure}[!ht]
    \centering 
    \includegraphics[width=11cm]{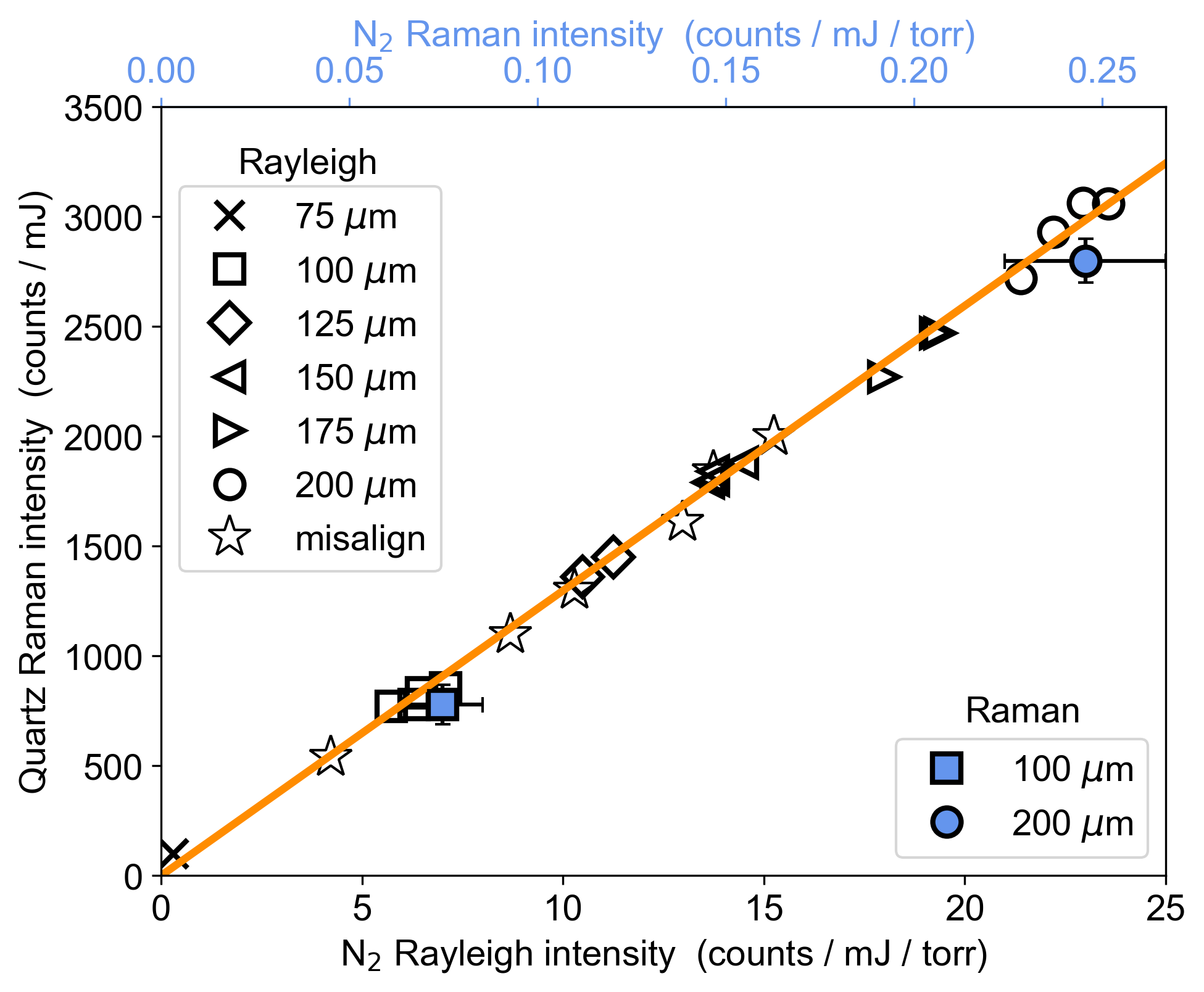} 
    \caption{\label{fig:calibration} Measured intensity of the 127 cm$^{-1}$ Raman line from quartz versus the Rayleigh and Raman intensities from nitrogen for different experimental configurations. The Raman axis (top) was scaled down by a factor of 94 with respect to the Rayleigh axis (bottom). While the intensities all decrease with slit width, the ratio stays constant. The star-shaped markers shows data obtained with a 150 $\mu$m slit while intentionally misaligning the collection branch relative to the probe laser beam.}
\end{figure}

Figure \ref{fig:calibration} compiles the measured intensity of the 127 cm$^{-1}$ quartz Raman line versus the Rayleigh and total Raman scattering signal from nitrogen for various experimental configurations. The Raman intensity axis (top) was scaled by a factor of 94 relative to the Rayleigh scattering axis (bottom), consistent with equation \ref{rayleigh-raman}.
This data were collected over the course of several weeks and the collection optics and crystal were realigned for each measurement pair. While the total photon count decreases with slit width, the ratio between the Rayleigh and quartz intensities, and the ratio between the Raman and quartz intensities stays constant. This confirms that the quartz Raman line can be used as a robust and reproducible light source for an absolute spectrometer calibration.  

\noindent
The linear fit between the quartz and Rayleigh intensities is given by
\begin{equation}
    N_{\rm{quartz}} = (129.7 \pm 0.1)\ torr\cdot \frac{N_R}{p}
\end{equation}
where $p$ is the gas pressure in torr used when measuring the Rayleigh scattering signal $N_R$. At $p=760$ torr the relation can be expressed as 
\begin{equation}
    N_{\rm{quartz}} = (0.1707 \pm 0.0002)\cdot N_R 
\end{equation} 
Since the relation between the Thomson scattering count and the Rayleigh count is
\begin{equation}
    \label{tsratio}
n_e = \frac{N_T}{N_R} \cdot \frac{1}{131} \cdot n_{gas} 
\end{equation}
where  n$_{gas} = 2.5\cdot 10^{19}\ cm^{-3}$\ ,
\begin{equation}
\label{sofiya}
n_e = \frac{N_T}{N_{\rm{quartz}}} \cdot (3.258 \pm 0.003) \cdot {10^{16}} cm^{-3}\ .
\end{equation} 
The good agreement between the Rayleigh and the Raman calibration despite more than three orders of magnitude difference in pixel counts confirms that the camera response is linear over the dynamic range used. 

When using the broad fused silica Raman spectrum for calibration, signal amplitude rather than integrated counts can be used. Comparing fused silica Raman amplitude to Rayleigh integrated intensity is appropriate since the fused silica spectrum is not instrument broadened.
At 540 nm the Raman signal of fused silica was measured to be (3.3 $\pm$ 0.3) $\cdot 10^4$ counts/nm. At identical experimental conditions the measured integrated Rayleigh scattering signal in nitrogen at one standard atmosphere was (2.7$\pm$0.1)$\cdot 10^{4}$ counts. In combination with equation \ref{tsratio} the fused silica cross calibration factor is
\begin{equation}
    n_e = \frac{N_T}{N_{\rm{SiO_2}}} \cdot (2.3 \pm 0.2) \cdot {10^{16}}\  \frac{cm^{-3}}{nm}\ ,
\end{equation} 
where $N_{\rm{SiO_2}}$ is measured in counts/nm.
Table \ref{table:summary} summarizes all cross-calibration factors between silica and nitrogen.
\begin{table}[htbp]
\centering
\caption{\label{table:summary} Overview of the cross calibration factors at room temperature.}
\smallskip
\begin{tabular}{|l|c|c|}
\hline
 & Thomson & N$_2$ Rayleigh \\
\hline\hline
 & & \\
N$_2$ Rayleigh  & $\frac{N_T}{N_R} = 131\cdot \frac{n_e}{n_{gas}}$ &   \\[3mm]
N$_2$ Raman & $\frac{N_T}{N_{Raman}}= 1.23\cdot 10^4 \frac{n_e}{n_{gas}}$ & $\frac{N_R}{N_{Raman}} = 93.7$\\
 & & \\
N$_2$ Stokes & $\frac{N_T}{N_{Stokes}} = 2.08\cdot 10^4 \frac{n_e}{n_{gas}}$ & $\frac{N_R}{N_{Stokes}} = 159$\\
 & & \\
N$_2$ $J=6\rightarrow 8$ & $\frac{N_T}{N_{6\rightarrow 8}} = 1.92\cdot 10^5 \frac{n_e}{n_{gas}}$ & $\frac{N_R}{N_{6\rightarrow 8}} = 1.47\cdot 10^3$ \\
 & & \\
Quartz 127 cm$^{-1}$ & $\frac{N_T}{N_{quartz}}=3.07\cdot 10^{-17}\ \frac{n_e}{cm^{-3}}$ & $\frac{N_R}{N_{quartz}} = 0.0771  \frac{p}{torr}$\\
 & & \\

SiO$_2$ at 540 nm & $\frac{N_T}{N_{SiO_2}}=4.3\cdot 10^{-17}\ \frac{nm}{cm^{-3}}\cdot n_e$ & $\frac{N_R}{N_{SiO_2}} = 0.011\ nm  \frac{p}{torr}$\\
  & & \\

\hline
\end{tabular}
\end{table}

It was important to investigate refraction at the crystal interface as significant crystal misalignment could displace the scattering volume. For that purpose, signal intensities were investigated as the crystal was intentionally misaligned by $\pm$~2.5$^o$ about each axis. The intensity of Raman scattered light did not change as a result of the misalignments. 
%
%
Similarly, the star-shaped markers in figure \ref{fig:calibration} represent data obtained with a 150 $\mu$m slit while intentionally misaligning the collection branch relative to the probe laser beam. 
While a mismatch decreases the measured intensity, the ratio between the quartz and the Rayleigh lines remains constant. 
%
These results are indicative of the scattering volume not being affected by refraction as long as the crystal alignment is kept within reasonable limits. 

\subsection{Temperature dependence of the quartz spectrum}
\label{sec:temperature}
When using a solid state calibration probe in the LAPD during plasma operation the plasma source can heat the crystal to temperatures close to 200$^o$C. The intensity of the 127 $cm^{-1}$ Raman line was therefore calibrated as a function of temperature. Temperature was controlled and monitored using ceramic heaters and a thermocouple, respectively.
Simultaneously, the variation of the Raman spectrum with temperature was measured and compared with theoretical predictions in order to develop a Raman scattering based temperature diagnostic for the crystal.

\begin{figure}[!ht]
    \centering 
    \includegraphics[width=15cm]{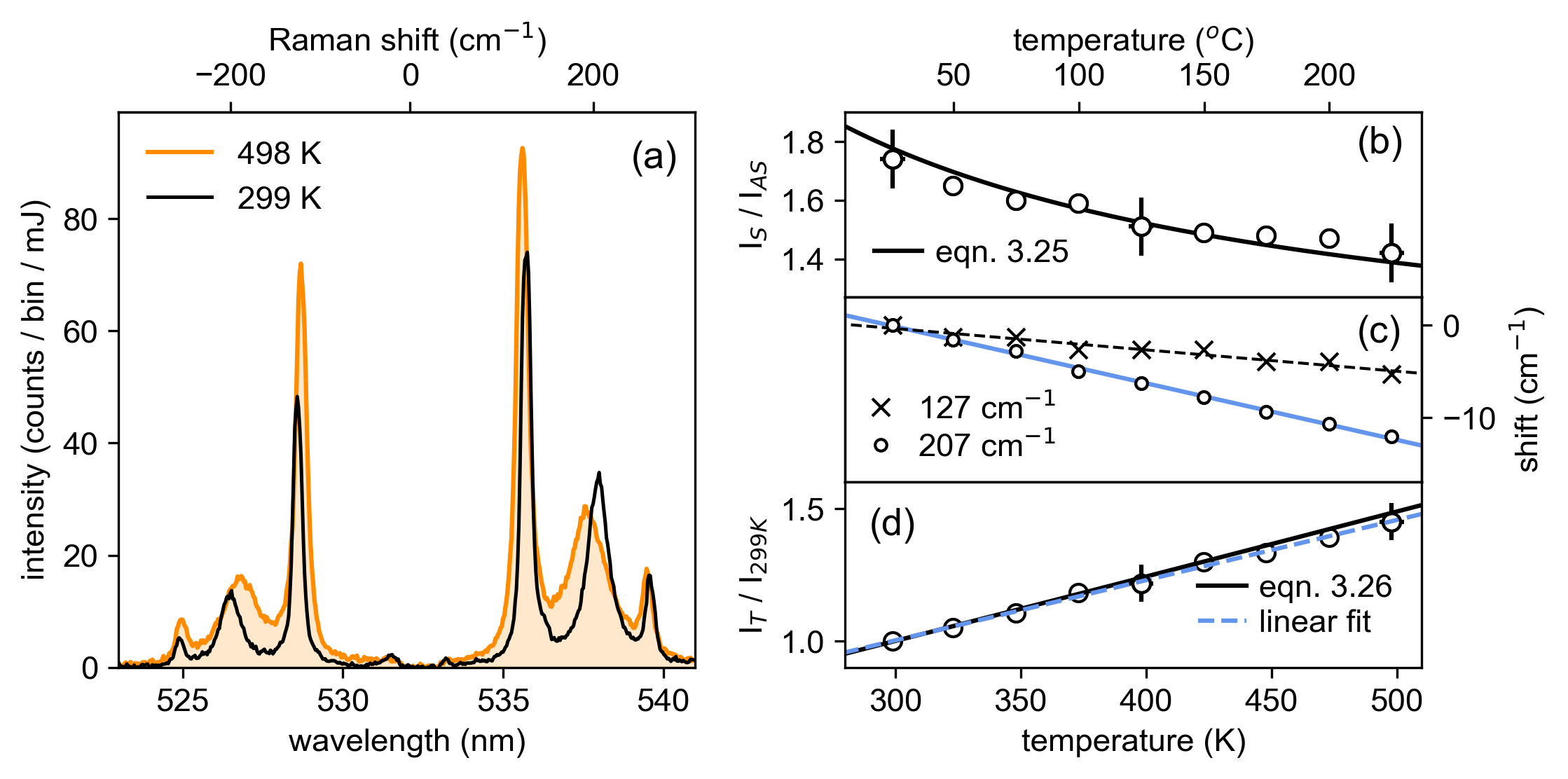}
    \caption{\label{fig:temperature} (a) The Raman spectra of quartz at room temperature (black) and at 225$^o$C (orange) show that line intensities and line width increase with temperature, while the Raman shift decreases.  (b) The measured Stokes versus anti-Stokes line intensity ratio agrees well with equation \ref{eqn:lineratio} for $n=3$. (c) The 127 cm$^{-1}$ and 207 cm$^{-1}$ Raman shifts decrease linearly with temperature and can be used as a more accurate temperature diagnostic. (d) The measured increase of the 127 cm$^{-1}$ line intensity with temperature agrees well with equation \ref{eqn:stokes}.}
\end{figure}

Figure \ref{fig:temperature}a compares the Raman spectra for room temperature (black curve) and for  225$^o$C (orange curve). 
Line intensities and widths generally increase with temperature, while the Raman shifts decrease. The 207 cm$^{-1}$ line shifts and broadens the most. Its Lorentzian line width increases from 0.84 nm (FWHM) at room temperature to 1.58 nm at 225$^o$C.
Both the line intensities of the Stokes and anti-Stokes lines increase with temperature. The anti-Stokes intensities increase faster than the Stokes intensities. The Stokes to anti-Stokes intensity ratio therefore decreases with temperature \cite{kip1990, mcgrane2014, gallardo2016}
\begin{equation}
    \label{eqn:lineratio}
    \frac{I_{S}}{I_{AS}} = \left(\frac{\nu_i - \tilde{\nu}}{\nu_i + \tilde{\nu}}\right)^n exp\left(\frac{h\tilde{\nu}}{k_B T}\right)\ ,
\end{equation}
where $\nu_i = c/\lambda_i = 5.64\cdot 10^{14}$ Hz is the frequency of the laser beam, $\tilde{\nu}=$ 127 cm$^{-1}\cdot c =3.83\cdot 10^{12}$ Hz is the transition frequency, and $c$ is the speed of light. The term in front of the exponential vanishes for Raman lines close to the laser wavelength. Previous experiments have found good agreement with various $n$. While Landsberg and Mandelstam \cite{landsberg1930} measured Stokes to anti-Stokes ratios in quartz that agreed best with the ordinary Boltzmann distribution function ($n=0$), spectra based on energy detection of the signals generally agree well with $n=4$, while $n=3$ is appropriate when spectra are recorded using photon counting in ICCDs \cite{tuschel2016}. Intensity ratios measured in this experiment shown in Fig. \ref{fig:temperature}b are consistent with $n=3$.
Line positions shift linearly with temperature (Fig. \ref{fig:temperature}b) consistent with other observations \cite{bauer2008}. The spectral shift of the 207 cm$^{-1}$ line is a particularly sensitive temperature diagnostic. For example, based on the fit in figure \ref{fig:temperature}c the crystal temperature can be determined from the measured line shift $\Delta \tilde{\nu}$ using $T = (297 - 16.3\ cm\cdot \Delta \tilde{\nu})\ K$ with an accuracy of 7 K.
The intensity of the Stokes lines increases with temperature \cite{hoebel1995} as 
\begin{equation}
    \label{eqn:stokes}
    I_{S}(T) \sim \left(\nu_i - \tilde{\nu}(T)\right)^4  \frac{1}{1- exp\left(\frac{-h\tilde{\nu}(T)}{k_B T}\right)\ }\ ,
\end{equation}
where the transition frequency $\tilde{\nu}$ of the Raman line also varies with temperature. This is consistent with the observations. Figure \ref{fig:temperature}c shows the measured ratio of the 127 cm$^{-1}$ Stokes line intensity at temperature $T$ and its intensity at T=299 K. The ratio increases by 50$\%$ for a temperature change of 200 K. A change in temperature of the order of the accuracy of the temperature measurement the intensity only changes by 1.5$\%$. Equation \ref{eqn:stokes} fits the observed intensity change very well. Over the temperature range of interest the ratio increases approximately linearly with temperature as
\begin{equation}
    \label{temp-fit}
    \frac{I_T}{I_{299\ K}} = 2.28\cdot 10^{-3} K^{-1} \cdot T + 0.320\ .
\end{equation}
The temperature dependent TS cross calibration relation for quartz is therefore
\begin{equation}
    n_e = \frac{N_T}{N_{\rm{quartz}}} \left( 7.4\cdot 10^{-3} \frac{T}{K} + 1.0\right)\cdot 10^{16}\ cm^{-3}\ .
\end{equation}

\section{Summary}
\label{sec:summary}
We have used Rayleigh and Raman scattering off nitrogen over a broad pressure range to cross calibrate the Raman spectra of crystalline (quartz) and amorphous fused silica.
The Raman spectrum of quartz shows three bright peaks at 127 cm$^{-1}$, 207 cm$^{-1}$, and 463 cm$^{-1}$. Since it is nearest to the laser line but beyond the notch, we used the 127 cm$^{-1}$ line as the calibration reference. 
Raman scattering off fused silica results in a broad band emission around 450 cm$^{-1}$.
%
Since the ratios between the measured Raman line intensities in quartz and the Raman and Rayleigh line intensities in nitrogen do not vary with the experimental parameters, the quartz Raman line can be used as a stable and reproducible signal for an absolute spectrometer calibration.
Measurements confirmed that the calibration obtained with Raman scattering off nitrogen fell in line with that of Rayleigh scattering to show the camera response is linear over the range used. 
Measured line intensities and line width increase with  temperature, while Raman shifts at 127 cm$^{-1}$ and 207 cm$^{-1}$ decrease linearly with temperature. 
The crystal temperature can be directly determined from the measured line intensity ratios and line shifts.
These measurements show that Raman scattering off quartz allows an accurate calibration of TS systems in plasma physics experiments, such as those to be performed in the LAPD, where a calibration via scattering off gases is not possible.

\acknowledgments
This work was supported by the Department of Energy under contract numbers DE-SC0019011 and DE-SC0021133.

\end{document}